\documentclass[aps,prl,showpacs, amsmath,amssymb,twocolumn]{revtex4}

\usepackage{graphicx}% Include figure files
\usepackage{dcolumn}% Align table columns on decimal point
\usepackage{bm}% bold math

\begin{document}

\title{Relaxation and flow in linearly sheared two-dimensional foams}
\author{Matthias E. M\"{o}bius, Gijs Katgert and Martin van Hecke}
\affiliation{Kamerlingh Onnes Lab, Universiteit Leiden, Postbus
9504, 2300 RA Leiden, The Netherlands}

\date{\today}

\begin{abstract}
We probe the relation between shear induced structural relaxation
and rheology in experiments on sheared two-dimensional foams. The
relaxation time $t_r$, which marks the crossover to diffusive bubble
motion, is found to scale non-trivially with the local strain rate
$\dot{\gamma}$ as $t_r \sim t_0^{0.34} \dot{\gamma}^{-0.66}$. Here
$t_0$ is the single-bubble relaxation time, which is four to seven
orders of magnitude smaller than the inverse strain rate
--- nevertheless, the flow is not quasistatic. The non-trivial
rheology of the foam is shown to be intimately linked to the
scaling of $t_r$, thus connecting macroscopic flow and microscopic
bubble motion.

\end{abstract}
\maketitle

%Foams, emulsions, colloidal glasses and granular media are soft
%glassy materials which jam into disordered, geometrically
%frustrated and densely packed structures \cite{???}.

Soft glassy materials, such as foams, emulsions, colloidal glasses
and granular media, exhibit highly complex flows. On the local
scale, rearrangements in these densely packed systems are erratic
due to geometric frustration. Globally, the relation between strain
rate $\dot{\gamma}$ and stress $\tau$ is generally non-linear, often
taking a Herschel-Bulkley form: $\tau=\tau_Y+c_1\dot{\gamma}^\beta$,
where $\tau_Y$ denotes the yield stress, and where the viscous
stress $\tau_V\equiv \tau-\tau_Y$ scales nontrivially with the
strain rate $\dot{\gamma}$ \cite{Larson,Hohler,bible}.

What is the connection between the global rheology and the
microscopic rearrangements of these materials?
%The macroscopic
%stress ultimately arises from rearrangements on the local scale.
For non-Brownian systems, the strain rate controls the dynamics,
therefore a fruitful strategy is to characterize and compare the
dependence of both stress and local relaxation events on the strain
rate. Recent experimental and simulation work on sheared foams has
uncovered rich dynamics on the bubble scale
\cite{Gopal1995,Gopal1999,Ono2003,Wang2,Katgert} that often shows a
non-trivial dependence of relaxation time scales on the strain rate
\cite{Gopal2003,Gopal1999,Ono2003}. However, except for ordered
foams \cite{Denkov2008}, a quantitative connection between these
time scales and the bulk rheology is still lacking.

 %Recent experimental and simulation work on sheared foams has
%uncovered rich dynamics on the bubble scale as well as non-trivial
%rheology, and various non-trivial scaling relations between
%relaxation time and strain rate have been observed, but without
%connecting these to the rheology
%\cite{wang,gopalprl2003,gopaljcis,onopre2003,wang,katgert}.

In this Letter we establish a direct and quantitative connection
between the strain rate dependencies of both the microscopic
relaxation time $t_r$, which embodies the decorrelation time of the
bubble motion, and the macroscopic viscous stress $\tau_V$ in
experiments on a sheared two-dimensional, disordered foam. %where the bubble motion is purely shear
%induced.
This approach is similar in spirit to recent work on glassy systems
where this connection has been probed \cite{Besseling, Varnik2006}

%by Besseling {\em et al.} for sheared colloidal glasses, where
%nontrivial scaling with strain rates of both relaxation time and
%stresses were observed, but where these two scaling laws could not
%be related easily \cite{besseling}.

\begin{figure}[tbhp]
\begin{center}
\includegraphics[width=3.2in]{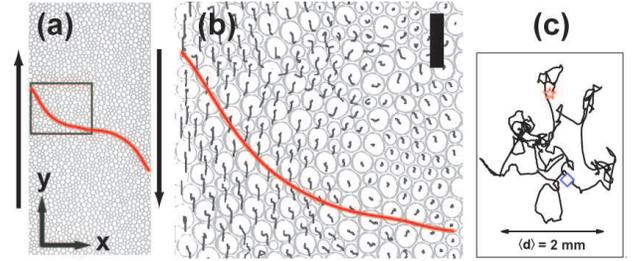}
%\vspace{-3.5cm}
\caption{\label{fig1} (color online) (a) Schematic top view of the
setup, with region of interest highlighted and typical flow
profile indicated by the red curve. (b) Zoom of the highlighted
region showing short time bubble trajectories. The width of the
black bar denotes the bin size $\Delta x=$ 1.6 mm. (c) Example of
a non-affine bubble trajectory over a period $\Delta t =145$ s,
for $\dot{\gamma}=0.03$ s$^{-1}$. The star and diamond symbol
denote the starting and endpoint, respectively.
%c) The non-affine mean square displacement
%$\langle \Delta s_y^2 \rangle$ versus local strain
%$\gamma=\dot{\gamma}\Delta t$ for different local shear rates
%ranging from $\dot{\gamma}=XXX$ to $xxx$. d) Same data as in (c),
%but plotted versus dimensionless
% time $\Delta t/ t_r(\dot{\gamma}) = \dot{\gamma} \mbox{whatever}$. The dotted line
%indicates the diffusive limit: $\langle \Delta s_y^2 \rangle = 2
%D_y \Delta t$, with $D_y=0.016XXerrorbar  \langle d \rangle^2
%/t_r$.}
}\vspace{-7mm}
\end{center}
\end{figure}

We measure the relaxation dynamics directly by imaging the bubble
motion, while the non-trivial rheology of this material is
determined from rheometric measurements \cite{Katgert,Denkov2005}.
%MAYBE motivate, we know this stuff is non trivial already (prl
%katgert).
%
%We establish a connection between the microscopic relaxation time
%$t_r$ and the macroscopic viscous stress $\tau_V$, by
%experimentally determining {\it both} the fluctuations in the
%bubble motion and the rheology of two-dimensional foams at steady
%shear \cite{Katgert,Janiaud}.
Figure \ref{fig1} shows examples of the bubble trajectories in a
top view of our experiment. The average flow profiles exhibit
shear bands in our experiment \cite{Katgert,Janiaud,
Wang,Langlois2008} and we employ these to study the statistics of
bubble displacements over a range of timescales and for local
strain rates spanning three decades.

First, we establish that, for a given local strain rate, the
probability distributions of bubble displacements exhibit fat
tails for short times, develop exponential tails for intermediate
times and finally become Gaussian. The occurrence of purely
exponential distributions at a sharply defined time allows us to
extract the relaxation time $t_r$.

Second, $t_r$ is not proportional with the inverse of the strain
rate, which would be the simplest relation consistent with
dimensional arguments, but instead exhibits a non-linear
relationship with the inverse strain rate: $t_r = (2.9 \pm 0.2)~
t_0^{1-\nu} \dot{\gamma}^{-\nu}$, where $\nu=0.66\pm 0.05$ and
$t_0$, the single bubble relaxation time \cite{Durian1995}, equals
$1.3\cdot 10^{-4}$ s.

Third, we find that the non-trivial scaling of the viscous stress
with strain rate \cite{Katgert,Langlois2008} can be directly
related to the scaling of the relaxation time: $\tau_V=(0.83 \pm
0.05)~ G_0 \gamma_r$, where $G_0$ is the static shear modulus and
the relaxation strain
is defined as $\gamma_r\equiv \dot{\gamma}t_r $. %\propto \dot{\gamma}^{1-\nu}
In addition we show that this connection is consistent with a
non-equilibrium Stokes-Einstein relation.

Our results firmly link strain rate, relaxation time, and stresses
in two-dimensional foams. We stress, however, that our scenario is
not particular to foams, and suggest to probe similar connections in
other sheared, non-Brownian systems, such as suspensions, granular
media, emulsions and microscopic models of these.

\begin{figure*}[tbp]
\begin{center}
\includegraphics[width=5.6in]{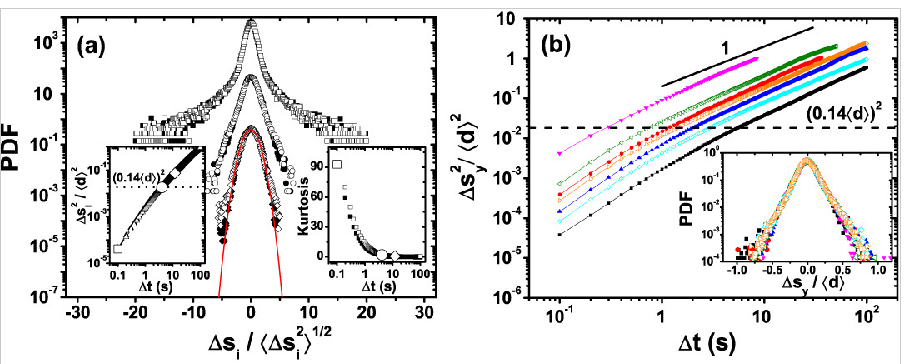}
\caption{\label{fig2} (color online) (a) Evolution of transverse
(full symbols) and longitudinal (open symbols) non-affine
displacement PDF's at $\dot{\gamma}=0.005 s^{-1}$:
($\blacksquare,\square$) $0.1$s; ($\bullet,\circ$) $4.1$s;
($\blacklozenge,\lozenge$) $12.1$s. The PDF's are shifted for
clarity and normalized by the width of the respective
distributions. The red curve is a Gaussian distribution. Left
inset: The corresponding longitudinal ($\vartriangle$) and
transverse ($\blacktriangle$) mean square displacements as a
function of time.
 Right inset: the Kurtosis $\langle \Delta s_i^4\rangle / \langle
\Delta s_i^2\rangle^2-3$ for the transverse (full symbols) and
longitudinal (open symbols) fluctuations. In both insets the three
symbols denote the time for which the PDF's have been plotted. (b)
Mean square displacements $\langle \Delta s_y^2\rangle$ as a
function of time at different local shear rates $\dot{\gamma}$ in
units of s$^{-1}$: $(\blacksquare)$ $0.002$; $(\lozenge)$ $0.009$;
$(\blacktriangle)$ $0.016$; $(\vartriangleright)$ $0.028$;
$(\bullet)$ $0.038$; $(\vartriangleleft)$ $0.11$;
$(\blacktriangledown)$ $0.29$. The horizontal line denotes the
Lindemann criterion $(0.14\langle d \rangle)^2$. The inset shows the
PDF's of the corresponding mean square displacements at the
Lindemann criterion.}\vspace{-7mm}
\end{center}
\end{figure*}

{\it Setup ---} The measurements are performed in a linear shear
cell (Fig.~1a) \cite{Katgert}, in which we create a bidisperse foam
layer with bubble diameters of $1.7$ and $2.7$ mm in a soapy
solution ($5\%$ volume fraction Dawn dishwashing liquid, $15\%$
glycerol and $80\%$ demineralized water, viscosity $\eta = 1.8~\cdot
10^{-3}$ Pa$\cdot$s, surface tension $\sigma = 28\cdot 10^{-3}$
N/m). The average bubble diameter $\langle d \rangle$ equals $2.0
\pm 0.1$ mm. The bubbles are trapped between the fluid layer and a
glass plate placed $2.2$ mm above the solution. Bubble coarsening is
negligible over experimental time scales and we do not observe
rupturing, coalescence or size separation of the bubbles. Local
shear rates encountered in this experiment range from $3\cdot
10^{-4}$ s$^{-1}$ to $0.3$ s$^{-1}$.

To induce shear, two rotating wheels of $39$ cm diameter are
partially immersed in a soap solution and spaced $7$ cm = 35
$\langle d \rangle$ apart. The rotation speed of the wheels controls
the driving velocity $v_0$ of the shearing wall which we vary
between $0.073$ and $2.3$ mm/s. The presence of the top glass plate
causes shear banding \cite{Wang, Katgert, Janiaud} and a locally
varying shear rate $\dot{\gamma}(x)$. We record the bubble motion by
imaging the monolayer from above at $10$ frames per second. Custom
tracking software allows us to determine the bubble positions $s_i$
to about $0.005 \langle d \rangle$. We focus on a central region
(Fig.~1a) where average flow transverse to the shear direction is
absent: $\langle v_x(x) \rangle = 0$.

{\it Mean Squared Displacements ---} In order to relate the
rearrangement rate of the bubbles and the local strain rate, we
divide the measurement area into bins of size $\Delta x \!=\! 1.6$
mm (Fig.~1b), and determine for each the local average flow $v(x)$
and shear rate $\dot{\gamma}\equiv
\partial v(x) / \partial x $. We then
determine the {\it non-affine} bubble tracks $\Delta s_i(\Delta
t)$ by subtracting the affine mean flow from the bubble
trajectories $s_i$: $\Delta s_i(\Delta t)\equiv s_i(t+\Delta
t)-s_i(t) - \langle v_i (x) \rangle  \Delta t $. An example of
an erratic bubble trajectory $\Delta s_i$ is shown in
Fig.~\ref{fig1}(c). We only consider
the mean square displacements on short time scales where
$\sqrt{\langle \Delta s_i^2 \rangle} \lesssim \langle d \rangle$,
so that Taylor dispersion is negligible.

In Fig.~\ref{fig2}(a), the probability distribution functions
(PDF's) of both the $x$ and $y$ component of $\Delta s$ are shown
for a given local strain rate $\dot{\gamma}$ at three different
times as indicated in the left inset of Fig.~\ref{fig2}(a). The
fluctuations of the bubble trajectories are isotropic,
% as evidenced
%by the collapse of the equal time PDF's of the transverse and
%longitudinal fluctuations and the mean square displacements.
%We
and we normalize the widths of all distributions by $\sqrt{\langle
\Delta s^2_i\rangle}$ to highlight the qualitative changes in these
PDF's with time. For short times, they exhibit ``fat tails'',
similar to the instantaneous velocity fluctuations observed in a
bubble raft \cite{Wang2}. Then, the PDF's develop exponential tails
and eventually become Gaussian in the long time, diffusive regime.
This behavior is also reflected in the evolution of the Kurtosis
(inset Fig~\ref{fig2}(a)).  We note that the time at which the PDF's
have exponential tails is well defined. This behavior should be
contrasted to sheared glasses, where the PDF's were found to have
exponential tails for a range of early times before crossing over to
Gaussian behavior at late times
\cite{Tanguy2006,Lemaitre2007,Chaudhuri2007}.

%Sheared granular system \cite{Radjai} and Lennard-Jones glasses
%\cite{Tanguy,Lemaitre,BertierPRL} display a similar evolution in the
%PDF's, except that exponential tails are observed for a range of
%early times before crossing over to diffusion.

In Fig.~\ref{fig2}(b) we show the longitudinal mean square
displacements for a range of local strain rates as function of time \cite{localfootnote}.
We find that the purely exponential PDF's occur when $\Delta s_y^2
\approx (0.14 \langle d \rangle)^2$. This also holds for the
transverse fluctuations \cite{Mobius}. Interestingly, this
corresponds to the Lindemann criterion for melting in atomic solids
and cage breaking in colloidal suspensions \cite{Besseling}. Here
the relaxation time $t_r$ marks the regime where the bubble motion
crosses over from super-diffusive to diffusive (see
Fig.~\ref{fig2}(b)). Even though $t_r$ can thus, in principle, be
defined from the PDF's or from this crossover, for simplicity, we
define $t_r$ as the time where $\Delta s_y^2 = (0.14 \langle d
\rangle)^2$. Note that the non-affine bubble displacements
between rearrangements are of similar magnitude (Fig.~\ref{fig1}(c)).

%applied shear rates: 0.002734869 0.008642185 0.086476547 0.273486865
%applied shear rates from velprof_y:  0.00217188   0.00713726    0.0431817    0.0643430 (slip)

\begin{figure}[tb]
\begin{center}
\includegraphics[width=3.1in]{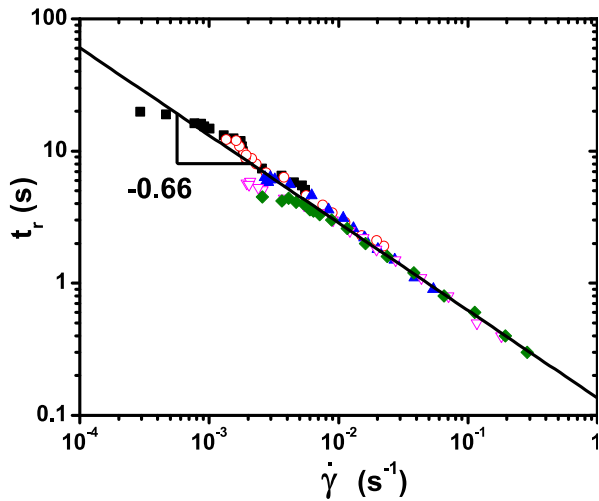}
\caption{\label{fig3} (color online) Relaxation time as a function
of local shear rate for different driving velocities:
($\blacksquare$) $v_0=0.076$ mm/s; ($\bigcirc$) $v_0=0.25$ mm/s;
($\blacktriangle$) $v_0=0.62$ mm/s; ($\triangledown$) $v_0=1.5$
mm/s; ($\blacklozenge$) $v_0=2.3$ mm/s. $t_r$ is defined as
$\langle \Delta s^2_y \rangle(\Delta t = t_r)=(0.14 \langle d
\rangle)^2$. The line is a fit to the data: $t_r=2.9 ~ t_0^{0.34}
\dot{\gamma}^{-0.66}$.}\vspace{-7mm}
\end{center}
\end{figure}

{\em Relaxation time ---} We now extract $t_r$ for a range of local
strain rates $\dot{\gamma}$ \cite{localfootnote}.
Since Brownian fluctuations do not play
a role, the rearrangements in the foam are entirely shear induced,
and {\it a priori} $\dot{\gamma}^{-1}$ would be a prime candidate
for setting the relaxation time. In contrast, we find that $t_r =
2.9 ~ t_0^{1-\nu} \dot{\gamma}^{-\nu}$, where $\nu \approx
0.66\pm0.05$, and $t_0$ is the characteristic relaxation time of
single bubbles $\eta \langle d \rangle / \sigma =1.3\cdot 10^{-4}$s
--- see Fig.~\ref{fig3}. This non-trivial scaling is our central
result, and it is far from obvious how it can be obtained, if at
all, from the local interactions of the bubbles.
%As we will show,
%this scaling implies that the flows are not quasistatic.

%Below we will directly link
%the scaling of $t_r$ to the non-trivial
%scaling of the viscous stress with strain rate.

%
%The non trivial scaling of $t_r$ REALLY IS COOL./ This non trivial
%scaling of relaxation time (DURIANS DWS DATA????)/ DENINS STEP
%STRAIN.... / many features of flow deeply connected to non trivial
%scaling.

\begin{figure}[b]
\begin{center}
\includegraphics[width=3.2in]{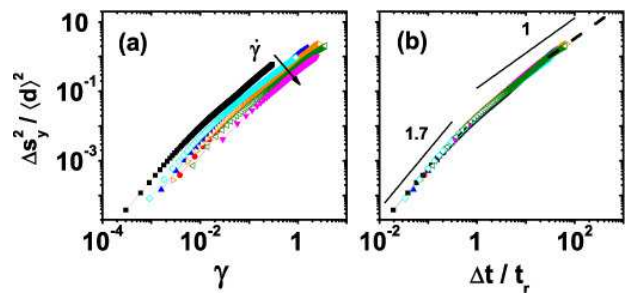}
\caption{\label{fig4} (color online) (a) The non-affine mean
square displacement $\langle \Delta s_y^2 \rangle$ versus local
strain $\gamma=\dot{\gamma}\Delta t$ for different local shear
rates (same data as Fig. \ref{fig2}(b)). (b) Same data as in (a),
but plotted versus dimensionless time $\Delta t/ t_r(\dot{\gamma})
\propto (\Delta t/t_0)^{0.34} \gamma^{0.66} $. The dotted line
indicates the diffusive limit: $\langle \Delta s_y^2 \rangle = 2
D_y \Delta t$, with $D_y=(0.017 \pm 0.002) \langle d \rangle^2
/t_r(\dot{\gamma})$.}\vspace{-7mm}
\end{center}
\end{figure}

{\em Data collapse for mean squared displacements ---} We now use
our findings to collapse the mean squared displacement curves onto a
single master curve. Fig.~\ref{fig4}(a) illustrates that the
system is not quasistatic, since the non-affine
mean square displacements do not simply scale with strain. In fact,
Fig.~\ref{fig4}(a) shows that for a given local average strain,
$\Delta s^2_y$ is larger in regions where the local strain rate is
smaller --- fluctuations thus increase for slower flows.

% Remarkably, $t_0$ is $4-7$
%orders of magnitude smaller than the inverse strain rates in our
%experiments -- nevertheless, this short timescale cannot be
%ignored and the system is not quasi-static.

Fig.~\ref{fig4}(b) shows that it is possible to obtain good data
collapse by rescaling the time axis with the relaxation time. We
then find a super-diffusive behavior with an initial slope $\approx
1.7$ for short times. %, which is similar to the early time mean-square
%displacement observed in granular shear flows \cite{Choi}.
Since large displacements due to bubble rearrangements are observed
even on the $0.1$ s time scale with which we image the foam, we do
not see a ballistic regime.
%Perhaps surprisingly, we do not see a ballistic regime at
%early times.
At later times the bubbles become diffusive and the
slope approaches $1$. From the collapse shown in Fig.~\ref{fig4}(b)
we deduce that the diffusion constant $D_y$ scales non-linearly with
the shear rate $\dot{\gamma}$ as $D_y=(0.017\pm 0.002) \langle d
\rangle/ t_r\propto \dot{\gamma}^\nu$. Similar results are found for
the transverse fluctuations \cite{Mobius}.

\begin{figure}[t]
\begin{center}
\includegraphics[width=3.2in]{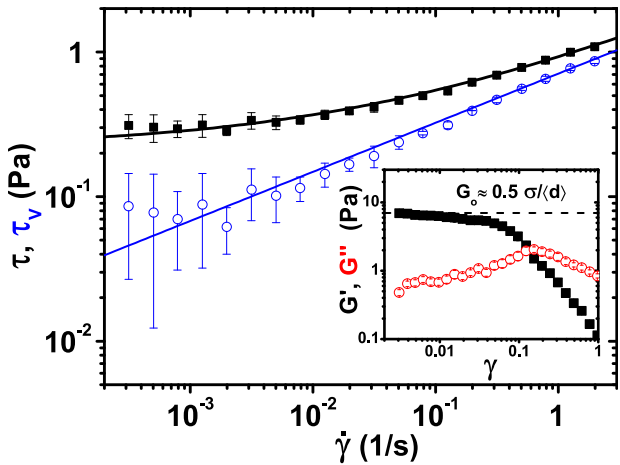}
\caption{\label{fig5} (color online) Foam rheology. The stress is
plotted as a function of shear rate measured in a Couette cell with
inner radius $5$ cm and outer radius $7$ cm. The disordered foam is
floating on the soap solution without the top plate and the
boundaries have grooves to prevent bubble slip. The shear rate at
the inner rotating wheel has been determined by image velocimetry.
($\blacksquare$) total stress. The Herschel-Bulkley exponent is
$\beta=0.36\pm0.05$. The curve through the data is $G_0\left
(\gamma_Y+0.83 ~ t_r \dot{\gamma}\right )$; ($\bigcirc$) viscous
stress $\tau_V\equiv \tau-\tau_Y$ where $\tau_Y=0.22\pm 0.2$ Pa. The
curve through the data is $0.83 ~t_r \dot{\gamma}$. Inset:
Oscillatory strain sweep at $0.1$ Hz. $G'$ approaches the static
shear modulus $G_0$ at low strain amplitude: $G_0\approx 7$ Pa and
the static yield strain $\gamma_Y=0.031$.}\vspace{-7mm}
\end{center}
\end{figure}

{\em Rheology ---}
%
%Our second main finding concerns the relation between the
%nontrivial relaxation timescale and the rheology of the foam.
We will now establish a surprising relation between relaxation times
and rheology. Fig.~\ref{fig5} illustrates that our 2D foams exhibit
a non-trivial rheology, with the shear stress $\tau$ equal to the
sum of a yield stress $\tau_Y$ and the rate dependent viscous stress
$\tau_V=c_1 \dot{\gamma}^{\beta}$ \cite{Katgert}. Remarkably, the
macroscopic rheological exponent $\beta$ and the microscopic
relaxation time exponent $\nu$ are related as $\beta \approx \nu-1$.

This connection between the viscous stress $\tau_V$ and $t_r$ can be
made intuitive by defining a relaxation strain $\gamma_r$ as the
product of relaxation time and strain rate. The relaxation strain is
thus increasing with shear rate, which is consistent with
observations by Rouyer et al. \cite{Rouyer}. To convert this
relaxation strain into a stress, we multiply it with the static
shear modulus $G_0$ which we measured to be ${\cal O} (
\sigma/\langle d \rangle)$ (see inset Fig.~\ref{fig5}) consistent
with Princen et al. \cite{Princen}. As Fig.~\ref{fig5} shows, $G_0
t_r \dot{\gamma}$ and the viscous stress $\tau_V$ are equal up to a
numerical factor of order one: $\tau_V = 0.83 ~G_0 t_r
\dot{\gamma}$. Hence the non-trivial global rheology of our foams,
and in particular the non-trivial power law scaling of the viscous
stress \cite{Katgert}, is deeply connected to the scaling of the
relaxation time $t_r$ with local strain rate.
%
%Similar results have been found in a recent simulation of a
%sheared glass \cite{Varnik2006}.

{\em Stokes-Einstein Relation ---} In previous numerical work on
sheared foams, a non-equilibrium Stokes-Einstein relation was found
to relate the product of diffusion constant, viscosity and bubble
size to an energy scale obtained from fluctuations in the elastic
energy \cite{Ono2002,Ono2003}. We show now that our connection
between relaxation time and viscous stress may also be viewed from
this perspective. Both the diffusion constant $D_y$ and the plastic
viscosity $\eta_p \equiv \tau_V / \dot{\gamma}$ are strain rate
dependent. However, the product of these quantities yields an energy
scale that is constant in the experimentally accessible range of
strain rates: $E=D_y\eta_p\langle d \rangle \approx 7\cdot 10^{-3}
~\sigma \langle d \rangle^2$, consistent with the simulation results at low strain rates \cite{Ono2002}. %\cite{ratefootnote}.
Since the elastic response of a single bubble can be modelled by a
spring with a spring constant $4\pi\sigma$, the energy scale $7\cdot
10^{-3} ~\sigma \langle d \rangle^2$ corresponds to compressing a
single bubble by $ \approx 0.04 \langle d \rangle$, which appears
reasonable \cite{Lacasse}.

{\em Discussion ---} Our central result is the scaling of the
relaxation time scales with strain rate: $t_r \approx t_0^{1-\nu}
\dot{\gamma}^{-\nu}$. Defining quasistatic flows as those for which
the precise bubble trajectories are independent of the flow rate,
the nontrivial scaling of $t_r$ and Fig.~\ref{fig4}a and \ref{fig4}b
clearly show that our flows are inconsistent with a quasistatic
picture, even though the global stresses approach rate independence
for slow flows. The microscopic scenario which emerges is that for
increasingly slow flows, the delicate balance between elastic and
viscous forces \cite{Liu1996} causes the (relative) bubble motion to
become increasingly concentrated in short bursts, such that the
contribution of viscous forces to the averaged stress vanishes ---
leading to rate independence of the stress \cite{Katgert}. This
scenario is consistent with the growth of $\Delta s_y^2$ at fixed
strain for decreasing $\dot{\gamma}$ (Fig.~\ref{fig4}a), and also
with the absence of a ballistic regime (Fig.~\ref{fig4}b): snapshots
of the displacement fields over a finite time interval reveal strong
spatially and temporally intermittent behavior, down to the shortest
timescale that we have probed.

Finally, it may seem counter-intuitive that our system is not
quasistatic, given that the single bubble relaxation time $t_0$
\cite{Durian1995} is several orders of magnitude smaller than the
typical inverse strain rate in our system. However, the fluid flow
is concentrated in the thin %(${\cal O} (100)$ nm)
liquid films separating the bubbles. The shear rate in these films
is therefore much closer to $1/t_0$, and the effective viscosity of
the foam is much higher than the viscosity of the fluid phase
\cite{Denkov2008,tomprivate}.

{\em Acknowledgements ---} The authors wish to thank J. Mesman for
technical assistance, and discussions with R. Besseling, T. Witten
and W. van Saarloos are gratefully acknowledged. GK and MM
acknowledge support from physics foundation FOM, and MvH
acknowledges support from NWO/VIDI.

%\nocite{*}


\begin{thebibliography}{99}
   % the 99 is as wide or wider than any bibliography labels.
\bibitem{Larson} R. G. Larson, {\it The Structure and Rheology of Complex Fluids} (Oxford University Press, New York, 1999).
\bibitem{Hohler} R. H\"ohler and S. Cohen Addad, J. Phys. Condens. Matter {\bf 17}, R1041 (2005). %cited
\bibitem{bible} D. Weaire and S. Hutzler, {\it The Physics of Foams} (Clarendon Press, Oxford, 1999).
\bibitem{Gopal1995} A. D. Gopal and D. J. Durian, Phys. Rev. Lett. {\bf 75}, 2610 (1995).
\bibitem{Gopal1999} A. D. Gopal and D. J. Durian,J. Colloid Interface Sci. {\bf 213}, 169 (1999).
\bibitem{Ono2003} I. K. Ono, S. Tewari, S. A. Langer and A. J. Liu, Phys. Rev. E {\bf 67}, 061503 (2003).
\bibitem{Wang2} Y. Wang, K. Krishan and M. Dennin, Phys. Rev. E {\bf 74}, 041405 (2006).
\bibitem{Katgert} G. Katgert, M. E. M\"{o}bius and M. van Hecke, Phys. Rev. Lett. {\bf 101}, 058301 (2008).
\bibitem{Gopal2003} A. D. Gopal and D. J. Durian, Phys. Rev. Lett. {\bf 91}, 188303 (2003).
\bibitem{Denkov2008} N.D. Denkov et al., Phys. Rev. Lett. {\bf 100}, 138301 (2008); S. Tcholakova et al., Phys. Rev. E {\bf 78}, 011405 (2008).
\bibitem{Besseling} R. Besseling, E. R. Weeks, A. B. Schofield, and W. C. K. Poon, Phys. Rev. Lett. {\bf 99}, 028301 (2007).
\bibitem{Varnik2006} F. Varnik and O. Henrich, Phys. Rev. B {\bf 73}, 174209 (2006).
\bibitem{Denkov2005} N. D. Denkov et al., Colloids Surf. A {\bf 263}, 129 (2005).
\bibitem{Janiaud} E. Janiaud, D. Weaire, and S. Hutzler, Phys. Rev. Lett. {\bf 97}, 038302 (2006).
\bibitem{Wang} Y. Wang, K. Krishan and M. Dennin, Phys. Rev. E {\bf 72}, 031401 (2006).
\bibitem{Langlois2008} V. J. Langlois, S. Hutzler and D. Weaire, Phys. Rev. E {\bf 78}, 021401 (2008).
\bibitem{Durian1995} D. Durian, Phys. Rev. Lett. {\bf 75}, 4780 (1995).
\bibitem{Tanguy2006} A. Tanguy, F. Leonforte and J. L. Barrat, Eur. Phys. J. E {\bf 20}, 355 (2006).
\bibitem{Lemaitre2007} A. Lema\^{\i}tre and C. Caroli, Phys. Rev. E {\bf 76}, 036104 (2007).
\bibitem{Chaudhuri2007} P. Chaudhuri, L. Berthier and W. Kob, Phys. Rev. Lett. {\bf 99}, 060604 (2007).
\bibitem{localfootnote} We have checked that our results only depend on the local strain rate by varying both the driving
velocity $v_0$ and location of the bin.
\bibitem{Mobius} M. E. M\"{o}bius, G. Katgert and M. van Hecke (in preparation).
\bibitem{Rouyer} F. Rouyer, S. Cohen-Addad, M. Vignes-Adler, and R. H\"ohler, Phys. Rev. E {\bf 67}, 021405 (2003).
\bibitem{Princen} H. M. Princen and A. D. Kiss, J. Colloid Interface Sci. {\bf 112}, 427 (1986).
\bibitem{Ono2002} I. K. Ono et al., Phys. Rev. Lett. {\bf 89}, 095703 (2002).
\bibitem{Lacasse} M.-D. Lacasse et al, Phys. Rev. Lett. {\bf 76}, 3448 (1996).
\bibitem{Liu1996} A. Liu at al., Phys. Rev. Lett. {\bf 76}, 3017
(1996).
%\bibitem{ratefootnote} We expect that for larger strain rates
%the energy scale will increase \cite{Ono2002}.
\bibitem{tomprivate} T. Witten, private communication.
%\bibitem{Schall} P. Schall, D.A. Weitz, and F. Spaepen, Science {\bf 318}, 1895 (2007).

%\bibitem{Wyss} H. Wyss et al., Phys. Rev. Lett. {\bf 98}, 238303 (2007).







%\bibitem{Radjai} F. Radjai and S. Roux, Phys. Rev. Lett. {\bf 89}, 064302 (2002).



%\bibitem{Choi} J. Choi, A. Kudrolli, R. R. Rosales, and M. Z. Bazant, Phys. Rev. Lett. {\bf 92}, 174301 (2004).








%\bibitem{Mason} T.G. Mason, J. Bibette, and D.A. Weitz, J. Colloid Interface Sci. {\bf 179}, 439 (1996).




%\bibitem{Ellenbroek} W.G. Ellenbroek, E. Somfai, M. van Hecke and W. can Saarloos, cond-mat/0604157 (2006).




%\bibitem{Breedveld} V. Breedveld et al., J. Chem. Phys. 116, 10529 (2002).

%\bibitem{Losert} W. Losert, L. Bocquet, T.C. Lubensky and J.C. Gollub, Phys. Rev. Lett., {\bf 85}, 1428 (2000); L. Bocquet, W. Losert, D. Schalk, T.C. Lubensky and J.C. Gollub, Phys. Rev. E,{\bf 65}, 011307 (2002).
%\bibitem{Mueth} D.M. Mueth, Phys.Rev.E, {\bf 67}, 011304 (2003).
%\bibitem{Debregeas} G. Debregeas, H. Tabuteau and J.M. diMeglio, Phys. Rev. Lett., {\bf }, 178305 (2001).



%\bibitem{Lauridsen2002} J. Lauridsen, M. Twardos, and M. Dennin, Phys. Rev. Lett., {\bf 89} 098303 (2002); E. Pratt and M. Dennin, Phys. Rev. E {\bf 67}, 051402 (2003).





%\bibitem{notemsd} In colloidal glasses, mean square displacements also change nature
%when this criterion is satisfied.

%\bibitem{viscosity} The ``plastic viscosity'' $\eta_p\equiv (\tau-\tau_Y)/\dot{\gamma}$ is usually only used
%for Bingham plastics, where it is a constant. Alternatively, one
%could use the differential viscosity $\eta_diff\equiv
%\partial\tau/\partial \dot{\gamma}.$

\end{thebibliography}
\end{document}